%% file: paper.tex
\documentclass[letterpaper,twocolumn,10pt]{article}
\usepackage{graphicx}
\usepackage{usenix,epsfig,bytefield}
\usepackage{ifthen}
\usepackage{xcolor}
\usepackage{booktabs}
\usepackage{color}
\usepackage{colortbl}
\usepackage{float}                           
\usepackage{bigdelim}

\usepackage[hyphens]{url}
\usepackage[colorlinks,citecolor=blue,breaklinks=true]{hyperref}%
\usepackage{svg}
\usepackage{lipsum}
\usepackage{amsmath}
\usepackage{amssymb}
\usepackage{amsthm}

\begin{document}

\newcommand\hmm[1]{\ifnum\ifhmode\spacefactor\else2000\fi>1000 \uppercase{#1}\else#1\fi}

\newcommand{\System}{QMDB: Quick Merkle Database}

\newcommand{\QMDBOverNOMTSpeedupHeading}{8X}
\newcommand{\QMDBOverNOMTSpeedup}{$8\times$}
\newcommand{\QMDBOverRocksDBSpeedupHeading}{6X}
\newcommand{\QMDBOverRocksDBSpeedup}{$6 \times$}
\newcommand{\QMDBMaxUpdatesPerSec}{2 million}
\newcommand{\QMDBMaxKeys}{15 billion}
\newcommand{\QMDBDRAMOverheadPerEntry}{15}
\newcommand{\RocksDBDRAMOverheadPerEntry}{1}
\newcommand{\QMDBSSDOverheadPerEntry}{210}
\newcommand{\RocksDBSSDOverheadPerEntry}{96}
\newcommand{\QMDBSixBilUpdatesPerSec}{601K}

\newcommand{\showComments}{yes}
\newcommand{\note}[2]{
    \ifthenelse{\equal{\showComments}{yes}}{\textcolor{#1}{#2}}{}
}
\newcommand{\nt}[1]{\note{blue}{Note: #1}}
\newcommand{\step}[1]{\raisebox{.5pt}{\textcircled{\raisebox{-.9pt} {#1}}}}

\widowpenalty10000
\clubpenalty10000

\date{}

\title{\Large \bf \System{}}

\author{
\normalfont{Isaac Zhang \hspace{0.7em} Ryan Zarick \hspace{0.7em} Daniel Wong \hspace{0.7em} Thomas Kim \hspace{0.7em} Bryan Pellegrino} \\ 
\vspace{0.4em} 
\normalfont{Mignon Li \hspace{0.7em} Kelvin Wong} 
\\
LayerZero Labs
}
\maketitle

\let\thefootnote\relax\footnotetext{\url{https://github.com/LayerZero-Labs/qmdb}}
\let\thefootnote\relax\footnotetext{Copyright \copyright{~2024 LayerZero Labs Ltd. All rights reserved.}}

\input{abstract}
\input{intro}

\input{background}
\input{design}
\input{evaluation}
\input{discussion}
\input{conclusion}

{\footnotesize
\bibliographystyle{acm}
\bibliography{bib}}


\end{document}

%% file: abstract.tex
\begin{abstract}
Quick Merkle Database (QMDB) addresses longstanding bottlenecks in blockchain state management by integrating key-value (KV) and Merkle tree storage into a single unified architecture. QMDB delivers a significant throughput improvement over existing architectures, achieving up to \QMDBOverRocksDBSpeedup{} over the widely used RocksDB and \QMDBOverNOMTSpeedup{} over NOMT, a leading verifiable database. Its novel append-only twig-based design enables one SSD read per state access, $O(1)$ IOs for updates, and in-memory Merkleization on a memory footprint as small as 2.3 bytes per entry enabling it to run on even modest consumer-grade PCs. QMDB scales seamlessly across both commodity and enterprise hardware, achieving up to 2.28 million state updates per second. This performance enables support for 1 million token transfers per second (TPS), marking QMDB as the first solution achieving such a milestone. QMDB has been benchmarked with workloads exceeding 15 billion entries (10× Ethereum’s 2024 state) and has proven the capacity to scale to 280 billion entries on a single server. Furthermore, QMDB introduces historical proofs, unlocking the ability to query its blockchain’s historical state at the latest block. QMDB not only meets the demands of current blockchains but also provides a robust foundation for building scalable, efficient, and verifiable decentralized applications across diverse use cases.
\end{abstract}

%% file: intro.tex
\section{Introduction}
\label{sec:introduction}
Updating, managing, and proving world state are key bottlenecks facing the execution layer in modern blockchains.
Within the execution layer, the storage layer, in particular, has traditionally traded off performance (throughput) and decentralization (capital and infrastructure barriers to participation).
Blockchains typically implement state management using an Authenticated Data Structure (\emph{ADS}) such as a Merkle Patricia Trie (\emph{MPT}).
Unfortunately, typical MPT-based ADSes incur a high amount of write amplification (\emph{WA}) with many costly random writes for each state update, which requires storing the entire structure in DRAM to avoid getting bottlenecked by the SSD.
As a result, the performance and scaling of blockchains is I/O-bound, and the key to unlocking higher performance with larger datasets is to optimize the use of SSD IOPS more efficiently and reduce WA.

We present Quick Merkle Database (\emph{QMDB}), a resource-efficient SSD-optimized ADS with in-memory Merkleization that implements a superset of the app-level features of existing RocksDB-backed MPT ADSes with \QMDBOverRocksDBSpeedup{} throughput on large datasets.
Qmdb performs state reads with a single SSD read, state updates with O(1) IO, and performs Merkleization fully \emph{in-memory} with no SSD reads or writes. These operations are theoretically optimal regarding disk IO complexity. Additionally, QMDB has a DRAM footprint small enough to run on consumer-grade PCs.


Blockchain state storage is typically handled by an Authenticated Data Structure (\emph{ADS}) which acts as a proof layer (e.g. Merkle Patricia Trie (\emph{MPT})) in combination with a physical storage layer.
The proof layer efficiently generates inclusion and exclusion proofs against the world state, while the physical storage layer stores the actual world state keys and values.
In many existing blockchains, these layers are each stored in a separate general-purpose key-value store such as RocksDB, resulting in duplicated data and general inefficiency.
Storing a MPT ($O(\log N)$ insertion) in a general-purpose key-value store ($O(\log N)$ insertion) results in each state update incurring $O((\log N)^2)$ SSD IOs.

QMDB eliminates this inefficiency by unifying the world state and Merkle tree storage, persisting all state updates in an append-only log, and eliminating all SSD reads and writes from Merkleization.
By grouping updates into fixed-size immutable subtrees called \emph{twigs}, QMDB can Merkleize state updates without reading or writing any world state; this essentially \textit{compresses} the Merkle tree by several orders of magnitude, allowing it to be stored in a modest amount of DRAM. QMDB leverages typical blockchain workload characteristics to eliminate features commonly found in KVDBs—such as key iterations—thereby reducing performance bottlenecks. 




These optimizations enable QMDB to achieve \QMDBOverRocksDBSpeedup{} throughput compared to RocksDB, a general-purpose key-value database that does not perform Merkleization.
We also show that QMDB outperforms a prerelease version of NOMT, a state-of-the-art verifiable database, by up to \QMDBOverNOMTSpeedup{}.
We validate QMDB's scaling characteristics with experiments up to 15 billion entries (10X of Ethereum's 2024 state size) and show it scales on both consumer-grade and enterprise-grade hardware.

QMDB is a transformative improvement for blockchain developers, addressing today's storage challenges and unlocking new possibilities for blockchain applications.
In particular:
1) QMDB can serve massive workloads with the same amount of DRAM, allowing blockchains to handle more users and transactions;
2) Based on its low memory overhead per entry, QMDB can theoretically scale up to 280 billion entries on a single server, far exceeding any blockchain's requirements today; and
3) QMDB can scale down to consumer-grade hardware, decreasing barriers to participation and improving decentralization.

%% file: background.tex
\begin{figure*}
    \centering
    \includegraphics[width=0.98\linewidth]{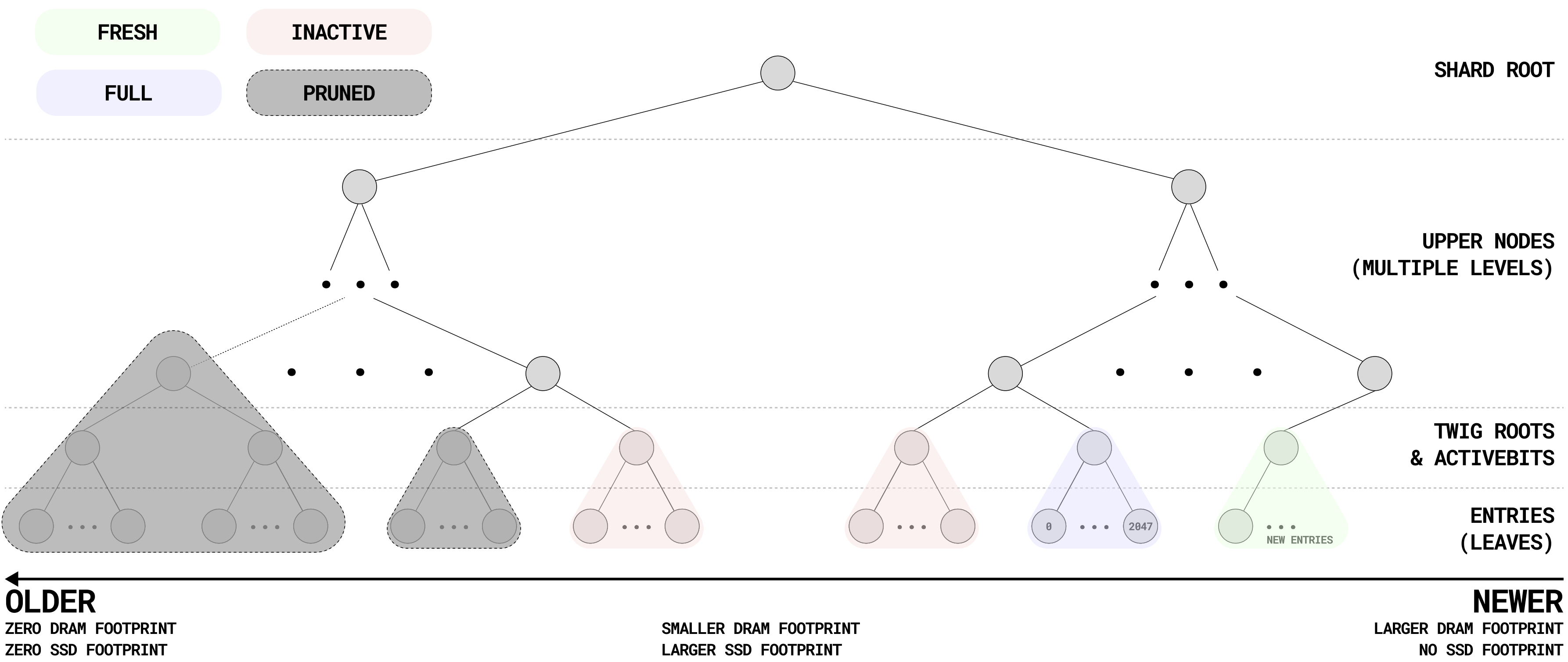}
    \caption{Entries are inserted sequentially into the leaves of the Fresh twig, and all leaves have the same depth. The twig eventually transitions into the Full state. As Entries are deleted, Full twigs become Inactive, then transition to Pruned. Upper nodes are recursively pruned after both of their children are pruned.}
    \label{fig:twig-lifecycle-example}
\end{figure*}
\section{Background}
\label{sec:background}

We explain the design of other verifiable databases and related data structures, including prior work reducing write amplification of verifiable databases~\cite{raju2018mlsm,li2024lvmt}.



\vspace{0.5em}
\noindent\textbf{MPTs} combine the efficient proof generation of the Merkle tree with the fast lookups of the Patricia trie and are a common choice for ADS on today's blockchains~\cite{wood2014ethereum}.
In a database of N items, updating a single state entry in an MPT has a time complexity of $O(\log(N))$~\cite{reth}.
However, MPT and other existing trie-based ADSes suffer from large proofs and a dependency on the client having a large amount of physical memory to avoid excessive random SSD reads.
At the same time, MPTs are not suitable for storage on flash storage, as the randomly distributed update-heavy workload results in high WA.
To top it off, the worst-case size for inclusion and exclusion proofs can be quite large.
These factors result in Merkleization becoming a significant bottleneck that limits the overall throughput of the execution layer and the blockchain.

\vspace{0.5em}
\noindent\textbf{AVL tree} based ADSes are popular alternatives to MPTs, as they achieve faster updates, lookups, and proof generation due to the self-balancing AVL tree. The AVL tree is path-dependent, unlike the MPT, meaning its state root is influenced by the specific sequence of state change actions.
AVL trees provide a marginal performance increase over MPTs in the average case, but still suffer from O(log N) tree nodes modifications per state update.

\vspace{0.5em}
\noindent\textbf{LVMT} $~\cite{li2024lvmt}$ proposes a layered storage model to reduce the space and complexity of maintaining authenticated blockchain states. To achieve constant-time Merkle root updates, it employs vector commitments and introduces a novel data structure to minimize the costly multiplications involved in vector commitment computations. In addition to the append-only Merkle tree, it incorporates an Addressable Merkle Tree (AMT) to support exclusion and latest-value proofs. However, maintaining an additional AMT with vector commitments increases resource overhead. Moreover, LVMT focuses on ADS design rather than providing a fully unified, verifiable database for end-to-end storage.

\vspace{0.5em}
\noindent\textbf{MoltDB}~\cite{liang2024moltdb} improves on existing two-layer MPT designs by segregating states by recency and coupling that with a compaction process. It reduces I/O and shows increased throughput of 30\% over Geth.

\vspace{0.5em}
\noindent\textbf{NOMT} is a state-of-the-art ADS that uses a flash-optimized layout for a binary Merkle tree with compressed metadata, overcoming some limitations of existing MPT-based ADS implementations. NOMT implements an array of improvements including tree arity, flash native layout, a write-ahead log, and caching. 
This design results in better performance than existing solutions and has garnered interest in the space. However, NOMT remains an implementation-level optimization of MPT, offering only constant-factor reductions in disk I/O. It still faces inherent asymptotic limitations and write amplification issues. Additionally, it is affected by the key sparsity problem commonly observed in trie-based structures.

\vspace{0.5em}
\noindent\textbf{Merkle Mountain Range (MMR)~\cite{todd_mmr_2016}} enable compact inclusion proofs and are append-only, which makes the IO pattern for updating state conducive to efficient usage of SSD IOPS.
Each MMR is a list of Merkle subtrees (\emph{peaks}), and peaks of equal size are merged as new records are appended.

MMRs are not suitable for live state management, as they cannot natively handle deletes, updates, lookups by key, and exclusion proof generation.
As a result, MMRs have generally found success in their use for historical data management~\cite{herodotus_mmr} where the key is just an index.

\vspace{0.5em}
\noindent\textbf{MoeingADS} ~\cite{moeingads} is another novel ADS design featuring an append-only Merkle tree and a memory-efficient representation for exclusion and latest-value proofs. Compared to LVMT, MoeingADS incurs even lower resource overhead while supporting the necessary proof features for ADS usability. However, its compaction and exclusion maintenance are performed in batches only after several blocks, impacting system performance and making it less suitable for stateless validation and Zero-Knowledge Proof-based verification. Additionally, MoeingADS lacks support for streaming transaction handling, a fine-grained pipelining system for higher throughput, and a low-memory indexer.

\vspace{0.5em}
\noindent\textbf{Acceleration of Merkle tree computation} has been an area of active research, with several proposed techniques such as caching~\cite{el2023towards,dahlberg2016efficient}, optimizing subtrees~\cite{ayyalasomayajula2023optimization}, and using specialized hardware~\cite{jeon2023hardware,deng2024accelerating}.
These improvements are orthogonal to QMDB and could be applied to QMDB to further improve its performance and efficiency.

\vspace{0.5em}
\noindent\textbf{Verifiable ledger databases} are systems that allow users to verify that a log is indeed append-only, of which blockchains are a subset.
A common approach to implementing a verifiable ledger database is deferred verification~\cite{yue2022glassdb,yang2020ledgerdb,antonopoulos2021sql}.
GlassDB~\cite{yue2022glassdb} uses a POS-tree (a Merkle tree variant) as an ADS for efficient proofs.
Amazon’s QLDB~\cite{amazon_qldb}, Azure’s SQLLedger~\cite{antonopoulos2021sql}, and Alibaba’s LedgerDB~\cite{yang2020ledgerdb} are commercially available verifiable databases that use Merkle trees (or variants) internally to provide transparency logs.
VeritasDB~\cite{sinha2018veritasdb} uses trusted hardware (SGX) to aid verification.
The key difference between these databases and QMDB is that QMDB is optimized for frequent state updates and real-time verification of the current state (as opposed to verification of historical logs and deferred verification).

%% file: design.tex
\begin{table*}[ht]
    \centering
\begin{tabular}{@{}rll@{}}
\toprule
\textbf{Field}          & \textbf{Description}                                                  & \textbf{Purpose}                       \\ \midrule
\(\text{Id}\)           & Unique identifier (e.g., nonce)                                       & Prove key inclusion                    \\
\(\text{Key}\)          & Application key                                              & Identify the key                       \\
\(\text{Value}\)        & Current state value the key                                           & Serve application logic                \\
\(\text{NextKey}\)      & Lexicographic successor of \texttt{Key}                & Prove key exclusion                    \\
\(\text{OldId}\)        & Id of the Entry previously containing \texttt{Key}     & Prove historical inclusion / exclusion \\
\(\text{OldNextKeyId}\) & Id of the Entry previously containing \texttt{NextKey} & Prove key deletion                     \\
\(\text{Version}\)      & Block height and transaction index                                    & Query state by block height            \\ \bottomrule
\end{tabular}
    \caption{\textbf{Fields in a QMDB entry.} ID and Version are 8 bytes. Key has up to $2^{8}$ bytes and Value can hold up to $2^{24}$ bytes.}
    \label{table:qmdb:entry}
\end{table*}

\section{QMDB Design}
\label{sec:design}
QMDB is architected as a binary Merkle tree illustrated in Figure~\ref{fig:twig-lifecycle-example}.
At the top is a single \emph{global root} that connects a set of \emph{shard roots}, each of which represents the subtree of the world state that is managed by an independent QMDB shard.
The shard root itself is connected to a set of \emph{upper nodes}, which, in turn, are connected to fixed-size subtrees called \emph{twigs}; each of these twigs has a root that stores the Merkle hash of the subtree and a bitmap called \text{ActiveBits} to track which entries are part of the most current world state. The twig root is determined by the sequence of entries, making it path-dependent. 
Entries (the twig's leaves) are append-only and immutable, making it unnecessary to read or write the entry root during Merkleization; this results in QMDB only ever reading/writing the global root, shard roots, upper nodes, and twig roots during Merkleization.
The twig essentially compresses the actual state keys and values into a single hash and bitmap, making the data required for Merkleization small enough to fit in a small amount of DRAM rather than being stored on SSD. 

In this section we begin by explaining the underlying storage primitives used to organize state data (Section~\ref{sec:entries-and-twigs}), followed by a discussion of the indexer in Section~\ref{sec:indexer}. 
In Section~\ref{sec:crud} we describe the high-level CRUD interface exported by QMDB to clients.
In Section~\ref{sec:proofs} we describe how the storage backend and indexer facilitate generation of state proofs, and discuss how these state proofs can be statelessly validated.
Finally, in Section~\ref{sec:parallelization} we explain how QMDB takes advantage of additional optimizations such as sharding and pipelining to scale throughput via improved parallelism.

\subsection{Entries and Twigs}
\label{sec:entries-and-twigs}

The \textbf{entry} (Table~\ref{table:qmdb:entry}) is the primitive data structure in QMDB, encapsulating key-value pairs with the metadata required for efficient proof generation.
Entries can be extended to support features such as historical state proof generation (Section~\ref{sec:proofs}).
QMDB keys entries by the \emph{hash} of the application-level key, resulting in improved load balancing via uniform key distribution across shards (Section~\ref{sec:parallelization})

\begin{table}[ht]
    \centering
    \begin{tabular}{@{}cccc@{}}
    \toprule
    \textbf{State} & \textbf{Description} & \textbf{Entries} & \textbf{Twig Root} \\ \midrule
    Fresh          & Entries $\leq 2047$  & DRAM             & DRAM               \\
    Full           & 2048 Entries         & SSD              & DRAM               \\
    Inactive       & 0 active Entries     & Deleted          & SSD                \\
    Pruned         & Subtree deleted      & Deleted          & Deleted            \\ \bottomrule
    \end{tabular}
    \caption{As twigs progress through their lifecycle, their footprint in DRAM gets smaller. An inactive twig has 99.9\% smaller memory footprint than a full twig.}
    \label{tab:twig-lifecycle}
\end{table}

\textbf{\emph{Twigs}} are subtrees within QMDB's Merkle Tree; each twig has a fixed depth, by extension a fixed number of entries stored in the leaf nodes of the same depth (2048 in our implementation).
A set of \emph{upper nodes} connects all twigs to a single shard root, with null nodes to represent uninitialized values; these upper nodes are immutable once all their descendant entries have been initialized.
In addition to the actual Merkle subtree, Twigs also store the Merkle hash of their root node and \emph{ActiveBits}, a bitmap that describes whether each contained entry contains state that has not been overwritten or deleted.
The twig essentially compresses the information required to Merkleize 2048 entries and their upper nodes ($\geq 256kb$) into a single 32-byte hash and a 256-byte bitmap (99.9\% compression).
This compression is the key to enabling fully in-memory Merkleization in QMDB.

Fresh twigs reside completely in DRAM, and entries are sequentially inserted into its leaf nodes.
Once a twig reaches 2048 entries, its contents are asynchronously flushed to SSD in a large sequential write and deleted from DRAM, maximizing SSD utilization and minimizing DRAM footprint.

Each twig follows a lifecycle of 4 states: Fresh, Full, Inactive, and Pruned (Table~\ref{tab:twig-lifecycle}).
An example of the layout of QMDB's state tree is presented in Figure~\ref{fig:twig-lifecycle-example}
There is exactly one fresh twig per shard, and entries are always appended to the fresh twig.
After all entries in the twig are marked inactive as a result of update and delete operations, the twig transitions into the inactive state before eventually being pruned and replaced by the Merkle hash of the root.
Upper nodes that contain only pruned twigs are recursively pruned, further reducing the memory footprint of QMDB; a dedicated garbage collection thread duplicates old valid entries into the fresh twig, reducing fragmentation and allowing larger subtrees to be pruned.
In theory, once the entire subtree originating at a child of the shard root is pruned, the root itself can be pruned to reduce the depth of the tree by one.

The grouping of entries into twigs reduces the DRAM footprint of QMDB to the degree that \emph{all nodes} affected by Merkleization can be stored in a small amount of DRAM.
In a hypothetical scenario with $2^{30}$ entries (approx. 1 billion), the system must keep at most $2^{19}$ ($\frac{2^{30}}{2048}$) 288-byte (32-byte twig root hash \& 2048-bit ActiveBits bitmap) full twigs, 1 fresh twig and $2^{19} - 1$ 32-byte (node hash) upper nodes totaling around 160 megabytes.
In practice, the majority of the $2^{19}$ twigs will be pruned, resulting in the average size being much smaller.

Inactive and Pruned twigs cannot be modified, and thus do not require further Merkleization.
Fresh and Full twigs must be Merkleized every time the ActiveBits bitmap is changed, and Fresh twigs must additionally be Merkleized every time an entry is added.
The upper nodes of the Merkle tree are recomputed on startup and are never persisted to SSD--this recomputation requires reading all twig hashes from SSD and performing 2 hashes per twig, and can be completed in a matter of milliseconds for the previous example of 1 billion entries.

QMDB stores an entry every time state is modified, making the state tree grow proportionally to the number of state modifications.
To combat this tree growth, a dedicated compaction worker periodically \emph{compacts} QMDB's state tree by removing and re-appending old entries to the fresh twig, accelerating the progression of the twig lifecycle and allowing more subtrees to be pruned. The compaction logic must be deterministic when used in a consensus system or for stateless validation. The current implementation ensures that the active entry ratio per shard remains above a predefined threshold, triggering compression during updates and insertions.

QMDB's Merkle proof size and proof generation complexity grow proportionally to $\log(U)$ of the number of state \emph{updates} ($U$) rather than the number of unique keys ($K$) due to its append-only nature. However, the ratio of $U$ to $K$ remains small enough that the order-of-magnitude improvement in Merkleization performance dominates the small additional cost.
Assuming 10,000 transactions per second and an average of 5 KV updates per transaction, the tree depth after one year will be at most 41 ($log2(10000 * 5 * 3600 * 24 * 365)$); however, in practice the actual depth will be much shallower due to pruning of overwritten subtrees and garbage collection.
In addition, ZK-proofs can be used to compress the proof witness data which drastically reduces proof verification cost, avoiding end-to-end bottlenecks in the proof size.

\subsection{Indexer}
\label{sec:indexer}
The \emph{indexer} maps the application-level keys to their respective entries, enabling QMDB's CRUD interface.
To support efficient insertion and deletion of entries (Section~\ref{sec:crud}), the indexer must support ordered key iteration.
The indexer can be freely swapped for different implementations depending on specific application needs, but we expect that QMDB's default in-memory indexer will meet the resource requirements of the majority of use cases.
This modularity potentially enables optimizations to increase the performance or memory efficiency of the indexer such as those found in systems such as SILT~\cite{silt} or MICA~\cite{mica}.

QMDB’s default indexer consumes approximately 15.4 bytes of DRAM per key and serves key lookups in-memory to minimize SSD I/Os. This efficiency is achieved by using only the 9 most significant bytes of each key, which slightly increases the likelihood of key collisions but strategically trades worst-case performance for reduced DRAM usage. Of these 9 bytes, the first 2 bytes serve as the sharding key for the indexer, leaving a 7-byte memory footprint for key storage. The remaining 8.4 bytes consist of a 6-byte SSD position offset and additional data structure overhead, which is amortized across all keys.
Using just 16 gigabytes of DRAM, the in-memory indexer can index more than 1 billion entries, making it suitable for a wide range of applications.
We chose the B-tree map as the basis for the underlying structure of QMDB's default indexer to take advantage of B-tree's high cache locality, low memory overhead, support for ordered key iteration, and graceful handling of key collisions.
We use fine-grained reader-writer locks (determined by the first two bytes of the key hash) to minimize contention when updating entries.

\subsection{CRUD interface}
\label{sec:crud}
QMDB exposes a CRUD (Create, Read, Update, Delete) interface, and in this section we provide a high-level overview of how each operation is implemented.
In all examples, we present the operation of the system when using the default in-memory indexer; other indexers may require more reads or writes to serve the same workload.
For each operation, we present an intuitive explanation followed by a more formal description along with a description of the SSD I/O required to synchronously handle the request.
All writes in QMDB are buffered in twigs (DRAM) and persisted to SSD in batches, so each SSD write is amortized across 2048 entries; to precisely express the cost of each operation, we refer to a \emph{entry write} as $\frac{1}{2048}$ of a single batched flush to SSD.
For brevity, we omit the Id, Version, and Value fields when describing new entries (see Table~\ref{table:qmdb:entry}), so an entry $E$ is defined as:
$$E = (Key,~NextKey,~OldId,~OldNextKeyId)$$

\vspace{0.5em}
\noindent\emph{\textbf{Read}} begins by querying the indexer for the file offset of the entry corresponding to a given key; this file offset is used to read the entry in a single SSD IO.

\vspace{0.5em}
\noindent\emph{\textbf{Update}} first \emph{reads} the most current entry for the updated key, then appends a new entry to the fresh twig.
More formally, if $E$ is the most current entry, the new entry $E'$ appended to the fresh twig derives its OldId and OldNextKeyId from $E$ as follows:
$$E' = (K,~E.nextKey,~E.Id,~E.OldNextKeyId)$$

Updating a key in QMDB incurs 1 SSD read and 1 \emph{entry write}.

\vspace{0.5em}
\noindent\emph{\textbf{Create}} intuitively involves appending one new entry and \emph{updating} one existing entry; the existing entry whose Key and NextKey define a range that coincides with the created key must be updated with a new NextKey.

This begins by first \emph{reading} the entry $E_p$ corresponding to the lexicographic predecessor ($prevKey$) to the created key $K$.
Note that $E_p$ must fulfill the condition $E_p.Key < K < E_p.nextKey$, as $K$ is not yet part of the current state.
Then, two new Entries are appended to the fresh Twig: 
$$E_K = (K,~E_p.nextKey,~E_p.Id,~E_n.Id)$$
$$E'_p = (prevKey,~K,~E_p.Id,~E_n.OldId)$$
The ActiveBit of $E_p$ is set to false (in memory), and the indexer is updated so that $prevKey$ points to the file offset of $E'_p$ and $K$ points to the file offset of $E_K$.

Creating a key in QMDB incurs 1 SSD read and 2 \emph{entry writes}.

\vspace{0.5em}
\noindent\emph{\textbf{Delete}} is implemented by first setting the activeBit to false for the most current entry corresponding to $K$, then \emph{updating} the entry for $prevKey$.
First, the entries $E_K$ and $E_p$ corresponding to the keys $K$ and $prevKey$ are read from SSD, and the ActiveBits for the twig containing $E_K$ is updated.
Next, a new entry for PrevKey is appended to the fresh twig:
$$E'_p = (prevKey,~E_K.nextKey,~E_p.Id,~E_K.OldNextKeyId)$$

Deleting a key in QMDB incurs 2 SSD reads and 1 \emph{entry write}.

\subsection{Proofs}
\label{sec:proofs}


The remainder of this section describes how each field of the QMDB entry enables the generation of various state proofs.
For illustrative purposes, we present proofs of the state corresponding to a key $K$ and the most current Merkle root $R$, and denote fields of an entry $E$ as $E.fieldName$.
All proofs are Merkle proofs and as a result can be statelessly verified.

\vspace{0.5em}
\noindent{\textbf{Inclusion}} is proved by presenting the Merkle proof $\pi$ for entry $E$ such that $E.Key = K$; this entry $E$ can be obtained after querying the corresponding file offset from the indexer.

\vspace{0.5em}
\noindent{\textbf{Exclusion}} is proved by presenting the inclusion proof of $E$ such that $E.Key < K < E.nextKey$.
The QMDB indexer supports efficient iteration by key, so $E$ can be located quickly by querying the lexicographic predecessor to $K$.




\vspace{0.5em}
\noindent{\textbf{Historical inclusion and exclusion}} at block height $H$ can be proven for a key $K$ by providing the inclusion proof of an entry such that $K$ is represented by this entry at the given version (block height).

QMDB uses \text{OldId} and \text{OldNextKeyId} to form a graph that enables the tracing of keys over time and space despite updates, deletions, and insertions.
\text{OldId} links the current entry to the last inactive entry with the same key and \text{OldNextKeyId} links to the entry previously referenced by \text{NextKey} (when the entry for \(\text{NextKey}\) is deleted).
When proving historical inclusion or exclusion, QMDB traverses the \text{OldId} pointer to move backwards in ``time'', and the \text{NextKey} and \text{OldNextKeyId} pointers to move to different parts of the key space at a given block height.





\vspace{0.5em}
\noindent{\textbf{Reconstruction of historical state}}
The graph structure defined by \text{OldId} and \text{OldNextKeyId} can also be used to reconstruct the Merkle tree and the world state at any block height.
The \text{Version} field tracks the block height and the transaction index where the entry was created, allowing the precise reconstruction of historical states at specific block heights.

\subsection{Parallelization}
\label{sec:parallelization}
\begin{figure}[ht]
    \centering
    \includegraphics[width=\columnwidth]{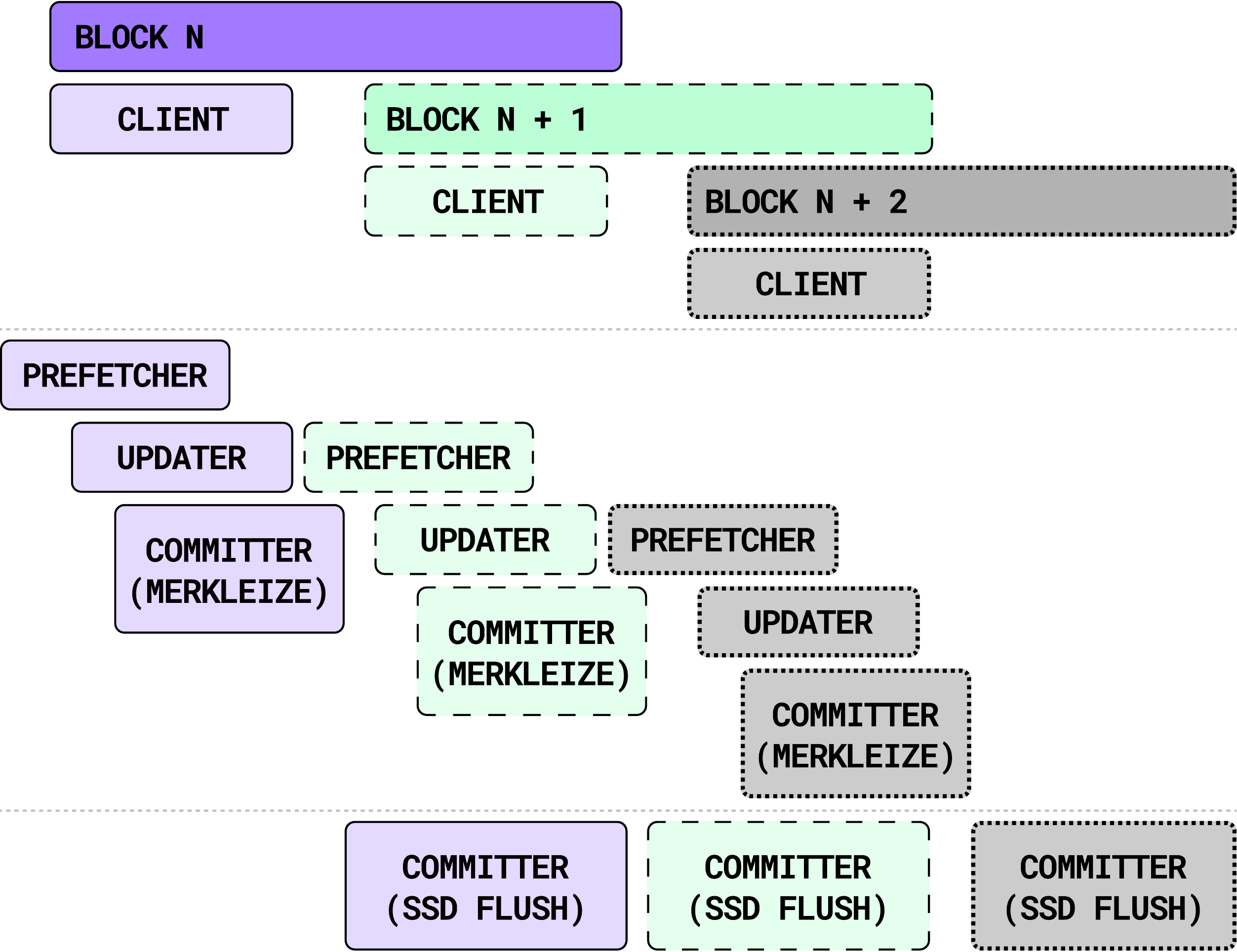}
    \caption{QMDB prefetches data (\emph{prefetcher}), performs the state transition (\emph{updater}), then commits the updated state to the Merkle tree and persistent storage (\emph{committer}).}
    \label{fig:pipelining}
\end{figure}
State updates are parallelized in QMDB through sharding and pipelining.

QMDB shards its key space into contiguous spans using the most significant bits—for example, the first 4 bits can create 16 shards—with \emph{boundary nodes} to define logical boundaries that prevent state modifications from crossing shard boundaries (i.e., PrevKey and NextKey will always fall within the same shard).
This sharding enables QMDB to better saturate underlying hardware resources and scale to bigger or multiple physical servers.

In addition, QMDB implements a three-stage pipeline (Prefetch-Update-Flush) to allow the transaction processing layer to better saturate QMDB itself.
For applications with relaxed synchronicity for state updates, QMDB is able to interleave computation across overlapping blocks.
This cross-block and intrablock parallelism allows QMDB to more fully saturate available CPU cycles and SSD IOPS.

Clients interact with QMDB by enqueueing key-value CRUD requests; updates are requested by writing the old Entry and new Value into the EntryCache directly, while deletions and insertions only require the key and new entry respectively.

The pipeline is illustrated in Figure~\ref{fig:pipelining}, and is managed by three workers: the fetcher, the updater, and the committer.
Each stage is shown in rectangles with solid lines, and the workers communicate via producer-consumer task queues in shared memory.
The fetcher reads relevant entries from SSD into the EntryCache in DRAM when necessary (Deletion and Insertion), while the updater appends new entries and updates the indexer.
Once the fetcher and updater finish processing a block, the committer asynchronously Merkleizes the updates and flushes the full twigs to persistent storage.

The QMDB pipeline has \emph{N+1 serializability}, which guarantees that state updates are visible in the next block.
This is implemented by enforcing that the prefetcher cannot run for block $N$ until the updater finishes processing block $N-1$.

%% file: evaluation.tex
\section{Evaluation}

In this section, we present a preliminary evaluation of the performance of QMDB and compare it to RocksDB and NOMT.
On a comparable workload and evaluation setup, QMDB achieves \QMDBOverRocksDBSpeedup{} higher updates per second than RocksDB and \QMDBOverNOMTSpeedup{} higher updates per second than NOMT.
When measuring the performance of QMDB, we generate 100,000 transactions per block--each creating 10 entries--and run the workload for 7000 blocks to create a total of 7 billion entries.
Periodically (every billion entries) we test the throughput and latency of reads, updates, deletions, and creations, and after all 7 billion entries are populated we measure transactions per second (TPS) using transactions consisting of 9 writes, 15 reads, 1 create, and 1 delete.

\subsection{\QMDBOverRocksDBSpeedupHeading{} more updates/s than key-value DBs}

\begin{figure}
    \centering
    \includegraphics[width=.8\linewidth]{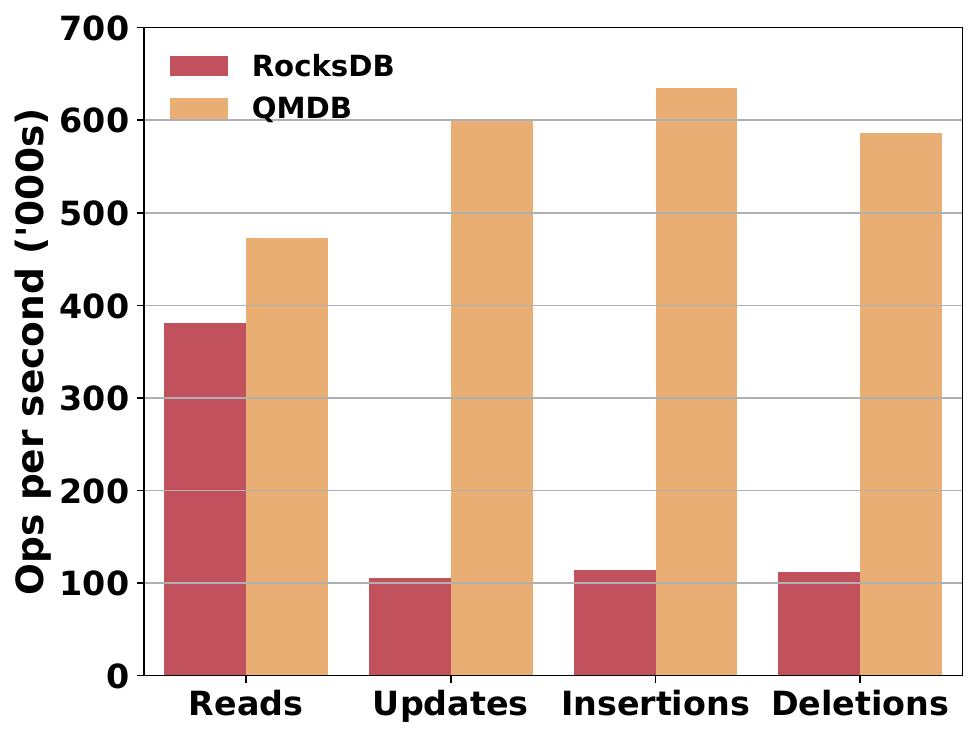}
    \caption{\textbf{QMDB shows a \QMDBOverRocksDBSpeedup{} increase in throughput over RocksDB.} QMDB is able to do \QMDBSixBilUpdatesPerSec{} updates/sec with 6 billion entries and demonstrates superior performance across all operation types. These results were obtained on an AWS c7gd.metal instance with 2 SSDs and 64 vCPUs.}
    \label{fig:eval:ops:c7g.metal}
\end{figure}

Figure~\ref{fig:eval:ops:c7g.metal} shows the throughput of QMDB compared to RocksDB (storing the application-level key-values with no Merkleization), demonstrating that QMDB delivers \QMDBOverRocksDBSpeedup{} more updates per second than RocksDB.
This speedup is in fact an underestimate of QMDB's advantage over RocksDB-based systems, given that all benchmarks compare QMDB with Merkleization to RocksDB without Merkleization.
We believe the primary factor driving this speedup to be QMDB trading off functionality unnecessary for blockchain workloads for extra throughput.
Examples of features and characteristics of RocksDB that are not required in blockchain workloads include efficient range/prefix queries and spatial locality of key-value pairs.


We caveat that our RocksDB evaluation is preliminary and could be better optimized, as our results were gathered on an unsharded RocksDB instance with default parameters.
We also tested RocksDB with the parameters recommended by the RocksDB wiki~\cite{rocksdb_wiki_setup_tuning} with direct I/O enabled for reads and compaction, but did not observe noticeably better performance.
We have also informally tested with MDBX but do not show those results here, as MDBX was significantly slower than RocksDB.

\subsection{Up to \QMDBOverNOMTSpeedupHeading{} throughput vs state-of-the-art}

For a more apples-to-apples comparison with a verifiable database that also performs Merkleization, we compared QMDB to NOMT~\cite{habermeier_introducing_nomt_2024}.
NOMT performs Merkleization and stores Merkleized state directly on SSD, and can be directly compared to QMDB in terms of functionality.
Both QMDB and NOMT aim to be drop-in replacements for general-purpose key-value stores like RocksDB, and aim to leverage the performance of NVMe SSDs. 

At the time of writing, both QMDB and NOMT are pre-release with significant optimizations in the pipeline for both systems, making a definitive comparison impossible at this point.
We used the version of NOMT from November 2024.
The steps we took to present a fair comparison include: evaluating QMDB and NOMT using their respective benchmark utilities, verifying the NOMT parameters with the authors~\cite{nomt_github_issue_611}, using the same hardware when evaluating each system, and normalizing the performance results against the workload.
Unfortunately, we were unable to eliminate all variability, as NOMT does not support client-level pipelining and the evaluated version of QMDB did not support direct IO or \texttt{io\_uring} (results for a newer version of QMDB with \texttt{io\_uring} and direct IO are shown in \S\ref{sec:qmdb-async-io}).

Table~\ref{table:eval:qmdb-nomt} shows the results of our evaluation, demonstrating a \QMDBOverNOMTSpeedup{} speedup in \emph{normalized updates per second} (transaction count multiplied by state updates per transaction).
NOMT's default workload is a 2-read-2-write transaction, whereas QMDB is evaluated with a 9-write-15-read-1-create-1-delete transaction (based on our own analysis of the operation composition of historical Ethereum transactions; data available upon request).
By normalizing the results based on the workload, we provide what we believe to be a fair representation of the comparative performance of these two systems.
The read latency was comparable ($30.7 \mu s$ for QMDB and $55.9 \mu s$ for NOMT) and close to the i3en.metal SSD read latency, which is in line with our expectations for both systems.

We believe this performance gap to be primarily driven by SSD write amplification, given that NOMT buffers in-place updates in a write-ahead log whereas QMDB's entries are immutable by design.
This results in persistent storage writes for potentially every state update and Merkleization for NOMT, compared to QMDB where a SSD write is only required every 2048 updates and zero SSD accesses are required for Merkleization.

We note that QMDB's performance relies on its indexer, which incurs some DRAM overhead. Compared to NOMT's overhead of 1--2 bytes per entry, QMDB incurs an additional 14 bytes per entry with its in-memory indexer and an additional 1--2 bytes per entry with its hybrid indexer). We consider this to be a reasonable trade-off given the \QMDBOverNOMTSpeedup{} increase in throughput, with QMDB's hybrid indexer still offering a speedup for DRAM-constrained setups.

\begin{table}[ht]
\centering
\begin{tabular}{@{}ccrc@{}}
\toprule
\multicolumn{4}{c}{\textbf{Normalized updates per second}}                                                                         \\ \midrule
\multicolumn{1}{c}{\textbf{\# Keys (M)}} & \multicolumn{1}{c}{\textbf{QMDB}} & \multicolumn{1}{c}{\textbf{NOMT}} & \textbf{Speedup} \\ \midrule
4 ($2^{22}$)                             & 614,948                           & 162,190                           & $4\times$               \\
256 ($2^{28}$)                           & 346,843                           & 42,277                            & $8\times$               \\
4096 ($2^{32}$)                          & 294,349                           & 37,057                            & $8\times$               \\ \bottomrule
\end{tabular}
\vspace{.5em}
\caption{\textbf{QMDB is up to \QMDBOverNOMTSpeedup{} faster than NOMT.}\\
Results are normalized by multiplying the transactions per second by the number of state updates per second.}
\label{table:eval:qmdb-nomt}
\end{table}


\subsection{Reaching 2M updates per second}
\label{sec:qmdb-async-io}
We show preliminary results indicating that QMDB can double its throughput and reach 2 million updates per second by incorporating asynchronous I/O (\texttt{io\_uring}) and direct I/O (\texttt{O\_DIRECT}), improving CPU efficiency and eliminating VFS-related overhead respectively.

Continuous advancements in SSD performance have resulted in modern \emph{consumer-grade} SSDs (e.g., Crucial T705, Samsung 980~\cite{habermeier_nomt_talk_2024}) being able to reach over 1 million IOPS with only one drive.
These high-IOPS SSDs are not yet available on AWS, so we approximate the performance in our preliminary experiments by using RAID0.

After populating QMDB with 14 billion entries, we measured 2.28 million updates/second on i8g.metal-24xl (6 SSDs) and 697 thousand updates/second on i8g.8xlarge (2 SSDs), which are promising early results.
2.28 million updates is sufficient to support over one million native token transfers per second (each transfer requiring two state updates).
QMDB's CPU utilization averages 77\% on the 32-core AWS i8g.8xlarge instance and 58\% on the 96-core AWS i8g.metal-24xl instance, indicating that with faster SSDs the bottleneck is no longer SSD IO but rather CPU and synchronization overheads.

We also evaluated NOMT with a lower capacity of 1 billion entries on the same instances (i8g.metal-24xl and i8g.8xlarge), and observed a maximum of 60,831 updates/second.
We acknowledge that comparing these numbers would not be fair given that NOMT is focused on supporting single-drive deployments, and RAID0 has different performance characteristics than a single SSD.
We plan a more comprehensive evaluation with a single high-performance SSD once we are able to secure a testbed with the necessary hardware.


 

\subsection{Scaling up \emph{and} down}
QMDB scales \emph{up} to huge datasets \emph{and down} to ultra-low minimum system requirements, enabling it to meet the needs of blockchains with the highest (performance-oriented) and lowest (most decentralized) node requirements.

\vspace{0.5em}
\noindent\textbf{\emph{Scaling up to hundreds of billions of entries.}}
The hybrid indexer trades off SSD capacity and system throughput to reduce the DRAM footprint of the QMDB indexing layer to just 2.3 bytes per entry, allowing servers with a high ratio of SSD capacity to DRAM capacity to scale to huge world states.
Table~\ref{table:eval:aws-datasize} shows the maximum theoretical number of entries that can be stored in QMDB running on various different AWS instances.
We calculate that the i3en.metal instance with high SSD capacity and a reasonable amount of DRAM could scale to \emph{280 billion} entries, far exceeding the needs of any existing production blockchain.
Due to the prohibitive amount of time necessary to populate hundreds or even tens of billions of keys, we only run experiments up to 15 billion entries and conservatively extrapolate the results.
The average DRAM overhead actually drops as more entries are inserted; 1 billion entries cost about 3 bytes of DRAM per entry, which drops to just 2.2 bytes per entry for 15 billion entries, indicating that the marginal DRAM overhead per additional entry is close to constant.

\begin{table*}[ht]
\centering
\begin{tabular}{lrrrrr}
\toprule
\multirow{2}{*}{\textbf{Instance type}} & \multirow{2}{*}{\textbf{DRAM (GiB)}} & \multirow{2}{*}{\textbf{SSD (TB)}} & \multicolumn{2}{l}{\textbf{Maximum entries (billions)*}} & \multirow{2}{*}{\textbf{Factor}} \\ 
                                        &                                     &                                    & \textbf{In-Memory}           & \textbf{Hybrid}          &                                  \\ \midrule
c7gd.metal                              & 128                                 & 3.8                                & 9.2                          & 18                       & 1.9                              \\
m7gd.metal                              & 256                                 & 3.8                                & 18.3                         & 18                       & 1.0                              \\
i3.metal                                & 512                                 & 15.2                               & 36.7                         & 71                       & 1.9                              \\
i8g.metal-24xl                          & 768                                 & 22.5                               & 55.0                         & 105                      & 1.9                              \\
i4i.metal                               & 1024                                & 30                                 & 73.3                         & 140                      & 1.9                              \\
i3en.metal                              & 768                                 & 60                                 & 55.0                         & 280                      & 5.1                              \\ \hline
\label{table:eval:aws-datasize}
\end{tabular}
\vspace{.5em}
\caption{\textbf{QMDB can scale to hundreds of billions of entries.} The hybrid indexer uses only 2--3 bytes of DRAM per entry. *This table shows extrapolated theoretical world state sizes for different hardware configurations, and compares the maximum entries stored using the in-memory indexer vs the hybrid indexer.}
\end{table*}


\vspace{0.5em}
\noindent\textbf{\emph{Scaling down to consumer-grade budget servers.}}
We built a low-cost Mini PC (parts totaling about US\$540 as of November 2024) to test QMDB under resource-constrained conditions.
The system featured an AMD R7-5825U (8C/16T) processor, 64 GiB DDR4 DRAM, and a TiPro7100 4 TB NVMe SSD rated at approximately 330K IOPS. Despite these modest specs, QMDB achieved tens of thousands of operations per second with billions of entries.
Using the in-memory indexer configuration, we were able to achieve 150,000 updates per second up to 1 billion entries, and stayed above 100,000 updates per second as we inserted up to 4 billion entries.
With the hybrid indexer, QMDB maintained 63,000 updates per second storing 15 billion entries.
These results highlight QMDB’s ability to operate on commodity hardware, improving decentralization by lowering the capital requirements and infrastructural barriers blockchain participation.

%% file: discussion.tex
\section{Discussion}
\label{sec:discussion}

\vspace{0.5em}
\noindent\emph{\textbf{Spatial locality}} is reduced in QMDB compared to general-purpose key-value stores such as RocksDB.
It is true that QMDB does not preserve temporal locality, given that keys that were originally inserted at similar times can become separated in QMDB if they are later updated.
However, this is not a disadvantage for blockchain workloads, given that blockchain infrastructure must assume worst-case workload characteristics 
to avoid exposing the blockchain to denial-of-service attacks in a Byzantine fault model.
This is unlike traditional computing workloads which can rely on locality for average-case performance.
In fact, most blockchains implement measures to uniformly distribute keys across storage with some exceptions (e.g., arrays in EVM); this already reduces or eliminates spatial locality.

\vspace{0.5em}
\noindent\emph{\textbf{Provable historical state}} enables new applications such as a TWAP (Time-Weighted Average Price)  aggregation at the tip of the blockchain with arbitrary time granularity.

\vspace{0.5em}
\noindent\emph{\textbf{Peer-to-peer syncing}} of state can be easily and efficiently implemented by sharing state at the twig granularity.
A downloaded twig accompanied by the inclusion proof of this twig against the global Merkle root can be inserted into the state tree independent of other twigs.

\vspace{0.5em}
\noindent\emph{\textbf{Memory-efficient indexers}} are useful for heavily resource-constrained use cases or for decentralization of blockchains with tens of billions of keys.
We implemented a memory-efficient SSD-optimized \emph{hybrid indexer} that uses only 2.3 bytes per key but requires one additional SSD read per lookup.
The hybrid indexer stores key-to-file offset mappings in immutable SSD-resident log-structured files and implements an overlay layer to manage entries in the SSD that have gone stale due to updates.
In addition, the hybrid indexer uses a DRAM cache of the spatial and temporal locality of the application workload.

\vspace{0.5em}
\noindent\emph{\textbf{State bloat}} is one of the many problems plaguing modern blockchains--as blockchains see growth in widespread adoption, world state is continuously growing to the point that it limits the ability of non-professional users to adequately run the validator software.
QMDB achieves a memory footprint that is an order of magnitude smaller than existing verifiable databases, and using the hybrid indexer can further reduce the memory footprint and decrease barriers to validator participation.

\vspace{0.5em}
\noindent\emph{\textbf{Recovery}} after failures (crash, blockchain reorganization) is done via replaying up to the last checkpoint and then trimming inactive entries.
The reference QMDB implementation intentionally omits specific reorg optimizations and leaves it up to individual blockchains, given the variation in consensus protocols between different chains. QMDB can be extended to support quick switches with an undo log, but in general QMDB expects blockchains to build a buffering layer on top of QMDB and only write finalized data (which is a similar approach to other verifiable databases).

\vspace{0.5em}
\noindent\emph{\textbf{Trusted Execution Environments}} (\emph{TEE}s) offer several security advantages to blockchains, and to the best of our knowledge QMDB is the first TEE-ready verifiable database.
Running a blockchain full node in a TEE (e.g., Intel SGX) protects the validator's private key from leaking, provides a secure endorsement that the state root was generated by a particular binary, guarantees peers that the validator is non-byzantine, and prevents censorship.

Current TEEs protect the integrity of CPU and DRAM, but cannot fully isolate persistent storage resources; QMDB protects its persistently stored data via AES-GCM~\cite{dworkin2007recommendation} encryption using keys dynamically derived from the virtual file offset to protect against copy attacks.

\vspace{0.5em}
\noindent\emph{\textbf{Zero-knowledge}} (ZK) proof generation for state transitions is increasingly seen as a crucial part of future blockchains, with one barrier to adoption being the long proof generation time.
The generation of ZK proofs can be parallelized per state commitment~\cite{roy2024introducingsp1} (e.g., each block can be proven individually and then chained together); thus, the degree of parallelization depends on the frequency of state root generation.
QMDB's high performance in-memory Merkleization is capable of computing a new state root \emph{per-transaction} if desired, enabling the maximum degree of parallelism for ZK proof generation.

%% file: conclusion.tex
\section{Conclusion}
\label{sec:conclusion}
QMDB represents a significant leap in blockchain state databases, providing an order of magnitude improvement in throughput over state-of-the-art systems in datasets $10\times$ larger than Ethereum at the time of writing.
Organizing and compressing state updates into append-only \emph{twigs}, QMDB is able to update and Merkleize world state with minimal write amplification, improving performance and reducing cost through efficient utilization of SSD IOPS.
The immutability of full twigs allows state to be compressed by more than 99.9\% for Merkleization, making it the first live-state management system capable of performing fully in-memory Merkleization with \emph{zero} disk IO on a consumer-grade machine.

We demonstrate that with these architectural innovations, QMDB can achieve up to \QMDBMaxUpdatesPerSec{} updates per second and scale to world states of \QMDBMaxKeys{} keys.
QMDB achieves lower minimum hardware requirements for all throughput benchmarks and world state sizes, democratizing blockchain networks by enabling affordable home-grade setups (US\$540) to participate in large blockchains.
At the same time, it provides substantial cost savings for large-scale operators due to its flash-heavy design that eliminates the need for large amounts of expensive and power-hungry DRAM.

QMDB implements many new features not present in other ADSes, such as historical state proofs, opening opportunities for a new class of applications not yet seen on the blockchain.
These features, together with order-of-magnitude advancements in performance and efficiency, establish QMDB as a breakthrough in scalable and verifiable databases.

\section{Acknowledgments}
We gratefully acknowledge the invaluable feedback and assistance provided by the many individuals and teams who helped refine our system. In particular, we thank Patrick O’Grady from Commonware for his expertise and guidance throughout the development of QMDB and the writing of this paper.  We thank the MoeingADS dev team for their review on throuhgput optimization. We also extend our gratitude to Yilong Li and Lei Yang from MegaETH, and Ye Zhang from Scroll, for their insightful review of our design and manuscript. Finally, we thank Robert Habermeier and the Thrum team for their support in conducting the NOMT benchmarks.